\documentclass[CEJP,DVI]{cej} 
\usepackage{layout}
\usepackage{amsmath}
\usepackage{textcomp}
\usepackage{hyperref}
\usepackage[T1]{fontenc}
\usepackage[english,polish]{babel}

\title{Charge-sensitive vibrational modes in the (EDT-TTF-OX)$_2$AsF$_6$ chiral molecular conductors}

\articletype{Topical Issue: Short Communication}

\author{Iwona~Olejniczak\inst{1}\email{olejniczak@ifmpan.poznan.pl},
        Arkadiusz~Fr\k{a}ckowiak\inst{1},
        Jacek~Matysiak\inst{1},\inst{2}
        Augustin~Madalan\inst{3},
        Flavia~Pop\inst{3},
        Narcis~Avarvari\inst{3}}

\institute{
     \inst{1} Institute of Molecular Physics Polish Academy of Sciences,\\
     Smoluchowskiego 17, 60-179 Pozna\'n, Poland
     \inst{2} Institute of Physics, Poznan University of Technology, \\ Nieszawska 13A,60-965 Pozna\'n, Poland
     \inst{3} Laboratoire Moltech Anjou, UMR 6200 CNRS, Universite d'Angers,\\
     UFR Sciences, B\^{a}t. K, 2 Bd. Lavoisier, 49045 Angers, France
          }

\abstract{Infrared and Raman spectra of the three chiral molecular conductors (EDT-TTF-OX)$_2$AsF$_6$, comprising two salts based on enantiopure
EDT-TTF-OX donor molecules and one based on their racemic mixture, have been measured as a function of temperature. In the frequency range of
the C=C stretching vibrations of EDT-TTF-OX, charge-sensitive modes are identified based on theoretical calculations for neutral and oxidized
EDT-TTF-OX using density functional theory (DFT) methods. The positions of the C=C stretching modes in both Raman and infrared spectra of the
(EDT-TTF-OX)$_2$AsF$_6$ materials are analyzed assuming linear relationship between the frequency and charge on the molecule. The charge density
on the EDT-TTF-OX donor molecule is estimated to be +0.5 in all the investigated materials and does not change with temperature. Therefore it is
suggested, that M-I transition observed in (EDT-TTF-OX)$_2$AsF$_6$ chiral molecular conductors at low temperature is not related to the charge
ordering mechanism.}


\begin{document}
\selectlanguage{english}
\maketitle

\begin{figure}
\includegraphics[width=0.3\textwidth]{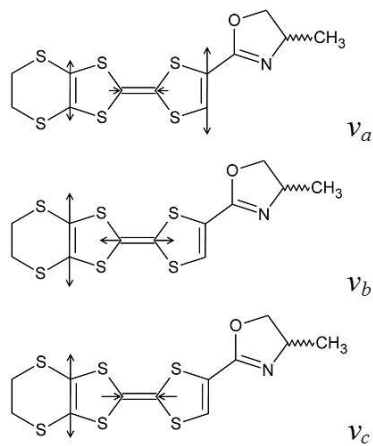}
\caption{Schematic mode pattern of the three fundamental C=C stretching modes of the EDT-TTF-OX molecule.\label{fig1}}
\end{figure}

Low-dimensional organic conductors based on bis(ethylenedithio)-tetrathiafulvalene molecule (abbreviated as BEDT-TTF) and its derivatives
continue to attract considerable attention because they are regarded as good model compounds to study the interplay between different phenomena
leading to broken symmetry ground states \cite{Ishiguro98}. Electronic parameters that are important for formation of different states in these
materials can be easily tuned by structural or/and chemical modifications. Their conducting properties are also significantly influenced by
disorder. This effect can be investigated in a convenient way if we introduce chirality as another functionality in the material
\cite{Avarvari09}. In case of molecular conductors, this can be done by using a chiral donor molecule or a chiral counterion. Of interest here
is the first complete series of metallic salts based on chiral ethylenedithio-tetrathiafulvalene derivatives with methyl-oxazoline heterocycle
as donor molecules \cite{Rethore05,Madalan10}, abbreviated here as EDT-TTF-OX (see Fig. \ref{fig1} for the structure). Three
(EDT-TTF-OX)$_2$AsF$_6$ salts have been synthesized, with donor molecules having (R), (S), and racemic (Rac) methyl-oxazoline heterocycle
attached to ethylenedithio-tetrathiafulvalene part. From the point of view of stoichiometry these materials have quarter-filled conduction band,
with the formal
charge +0.5$e$ per donor molecule.\\

The racemic (EDT-TTF-OX)$_2$AsF$_6$ (Rac) salt crystallizes in the triclinic system, space group $P\bar{1}$, with one donor molecule in general
position in the unit cell, and the oxazoline ring almost coplanar with the tetrathiafulvalene (TTF) core but disordered on two positions,
corresponding to both $R$ and $S$ enantiomers \cite{Rethore05,Madalan10}. Enantiopure (EDT-TTF-OX)$_2$AsF$_6$ (R) and (EDT-TTF-OX)$_2$AsF$_6$
(S) are isostructural and crystallize in the chiral space group $P$1 with two crystallographically independent donor molecules. All the three
materials have layered structure of the $\beta$-type according to classification developed for BEDT-TTF-based two-dimensional conductors, with
stacks of donor molecules along $b$, interacting along $a$ within the $ab$ conducting plane \cite{Rethore05,Madalan10}. Close inspection of the
dimerized donor arrangement and intrastack interaction energies \cite{Madalan10} suggests that the structure of the donor layer is characterized
by the $\beta$$^\prime$-sub-type, that often results in quasi-one-dimensional (Q1D) band structure and moves the material to an effective
half-filling \cite{Seo04}. Based on the results of both tight-binding calculations and conductivity measurements, the (EDT-TTF-OX)$_2$AsF$_6$
materials are Q1D metals with the highest conductivity direction along the stacking axis, as expected \cite{Rethore05,Madalan10}. Radical cation
salts of enantiopure donors are in principle less influenced by structural disorder than those of racemates where the two different enantiomers
are present in the crystal structure. In fact, the conductivity of the racemic (EDT-TTF-OX)$_2$AsF$_6$ (Rac) salt at room temperature is about 1
order of magnitude smaller than the conductivity of the two enantiopure materials \cite{Rethore05}. In case of all the (EDT-TTF-OX)$_2$AsF$_6$
salts, resistivity displays
a broad increase with lowering the temperature, leading to a localized regime below about 200 K.\\

One of the mechanisms that can drive this metal-insulator (M-I) transition is charge order (CO) originating in quarter-filled organic conductors
from the inter-site Coulomb repulsion \cite{Seo04,Yamamoto02}. Recently, both theoretical and experimental studies revealed that CO fluctuations
can participate in stabilization of the superconducting state \cite{Merino01,Merino03,Kaiser10}. Another possible mechanism of the M-I
transition in this case can be related to structural charge density wave (CDW) modulation \cite{Rethore05}. To answer the question about the
origin of the M-I transition in the (EDT-TTF-OX)$_2$AsF$_6$ materials we use vibrational spectroscopy that is a well-known sensitive probe of
charge distribution and bonding in the solid state. Investigations of charge localized on donor molecule in BEDT-TTF-based materials are focused
on the C=C stretching modes because their frequencies display strong dependence on charge on the molecule. The Raman-active $\nu_2$ and
infrared-active $\nu_{27}$ modes are most useful for estimation of charge \cite{Yamamoto05}. On the other hand, the totally symmetric $\nu_3$
mode is involved in the electron-molecular vibration (EMV) coupling \cite{Rice76}. Therefore it can serve as a sensitive probe of the local symmetry
breaking including dimerization.\\

In this work, variable temperature infrared and Raman spectra are used in order to characterize the M-I transition in the
(EDT-TTF-OX)$_2$AsF$_6$ chiral organic conductors. We focus our attention on the frequency range of the C=C stretching vibrations. Vibrational
features are assigned with the help of DFT calculations. In particular, we look for the signatures of CO at
low temperature. Our overall goal is to provide more information on structure-property relationship in this class of chiral organic conductors.\\

The three (EDT-TTF-OX)$_2$AsF$_6$ materials were synthesized and crystallized according to a procedure described previously \cite{Rethore05}.
Typical dimensions of samples used in infrared and Raman measurements were 0.4$\times$0.1$\times$0.1 mm$^3$. Single-crystal polarized
reflectance spectra in the frequency range 650 - 7500 cm$^{-1}$ were measured using a Bruker Equinox 55 FT-IR spectrometer equipped with a
Hyperion 1000 infrared microscope. The optical axes of the crystals were determined as those displaying the largest anisotropy at 300 K. Two
directions within the conducting plane were probed, the stacking $b$ direction and the $\perp$$b$ direction that is close to $a$ in the crystal
structure \cite{Rethore05}. Single crystalline Raman spectra were measured in a backward scattering geometry with a Raman LABRAM HR800
spectrometer, with a spectral resolution of 2 cm$^{-1}$. He-Ne ($\lambda$=632.8 nm) laser line was used with power reduced to about 0.1 mW to
avoid sample overheating. Both infrared reflectance and Raman spectra were measured at several temperatures between 10 and 300 K, and the
temperature was controlled with an Oxford Instruments continuous-flow cryostat. In order to assign vibrational features of EDT-TTF-OX,
theoretical calculations for isolated molecule were performed with Gaussian 03 \cite{Frisch03}, using the 6-31+G(d) basis set and the hybrid
Hartree-Fock density functional (B3LYP). $C1$ symmetry was identified at the level of theory that we used. On the basis of optimized structures,
vibrational frequencies as well as infrared intensities and Raman scattering activities were calculated. The frequencies computed with a quantum
harmonic oscillator approximation tend to be higher than experimental ones, so that in our study we used the scaling factor of 0.9614
\cite{Scott96}. In addition, both infrared and Raman spectra of neutral EDT-TTF-OX \cite{Rethore04} (powder sample, 300 K) and the racemic
(EDT-TTF-OX)$_2$Mo$_6$Cl$_{14}$ \cite {Rethore05a} (single crystal, 25 K) were measured in order to improve the assignment of the C=C
stretching modes.\\

\begin{figure}
\includegraphics[width=0.4\textwidth]{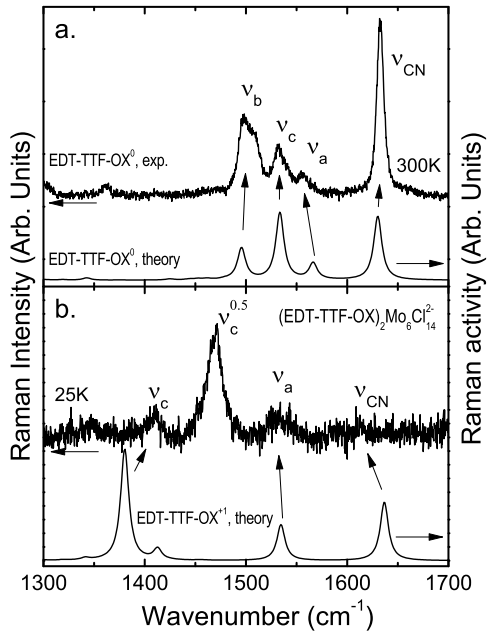}
\caption{Experimental Raman spectra of neutral EDT-TTF-OX (a) and (EDT-TTF-OX)$_2$Mo$_6$Cl$_{14}$ (b), together with the respective theoretical
Raman spectra of EDT-TTF-OX calculated for the neutral (a) and the molecule with the charge +1 (b). Note that the spectra are offset for
clarity.\label{fig2}}
\end{figure}

\begin{figure}
\includegraphics[width=0.4\textwidth]{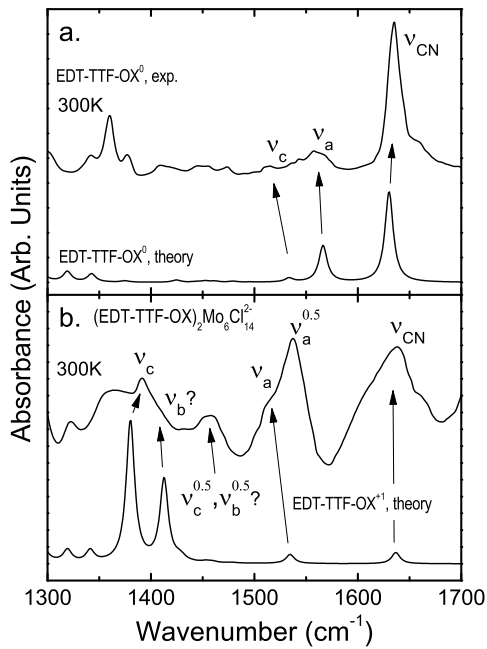}
\caption{Experimental room temperature infrared spectra of neutral EDT-TTF-OX (a) and (EDT-TTF-OX)$_2$Mo$_6$Cl$_{14}$ (b), together with the
respective theoretical infrared spectra of EDT-TTF-OX calculated for the neutral (a) and the molecule with the charge +1 (b). Note that the
spectra are offset for clarity.\label{fig3}}
\end{figure}

There are three C=C double bonds in the EDT-TTF-OX molecule, so that three fundamental C=C stretching modes are expected. Schematic pattern of
the modes based on our theoretical calculations is shown in Fig. \ref{fig1}. Here we assign the three C=C stretching modes of the EDT-TTF-OX
donor molecule as $\nu_a$ (in-phase stretching of the two side C=C bonds), $\nu_b$ (in-phase stretching of the bridge and side C=C bonds), and
$\nu_c$ (out-of-phase stretching of the bridge and side C=C bonds). In order to assign C=C stretching modes in our experimental spectra from the
point of view of the oxidation state of EDT-TTF-OX, we have performed theoretical calculations of vibrational modes for both the neutral
EDT-TTF-OX$^0$ and the radical cation EDT-TTF-OX$^{+1}$. Because of low symmetry, vibrational modes of EDT-TTF-OX are both infrared- and
Raman-active. In the frequency range 1300 - 1700 cm$^{-1}$, in theoretical Raman spectrum of EDT-TTF-OX$^0$ we observe four strong modes (Fig.
\ref{fig2}a). The highest frequency mode centered at 1631 cm$^{-1}$ is related to C$\equiv$N stretching vibration of the oxazoline ring
($\nu_{CN}$). The three described above C=C stretching modes are expected at 1566 ($\nu_a$), 1533 ($\nu_c$), and 1496 ($\nu_b$) cm$^{-1}$. We
can easily locate all these four vibrational features in the experimental Raman spectrum of neutral EDT-TTF-OX (upper curve in Fig. \ref{fig2}a)
at 1633 ($\nu_{CN}$), 1557 ($\nu_a$), 1532 ($\nu_c$), and 1499 ($\nu_b$) cm$^{-1}$. Theoretical infrared spectrum (Fig. \ref{fig3}a) displays
the same modes as the respective Raman spectrum, but with different band intensities.  These features are found in experimental infrared spectra
of EDT-TTF-OX$^0$
(upper curve in Fig. \ref{fig3}a) at 1636 cm$^{-1}$ ($\nu_{CN}$), 1556 cm$^{-1}$ ($\nu_a$) and 1514 cm$^{-1}$ $\nu_c$.\\

Now, let us closely examine these modes in experimental and theoretical Raman spectra of the EDT-TTF-OX$^{+1}$ cation (Fig. \ref{fig2}b).
Theoretical $\nu_{CN}$ mode is centered at 1636 cm$^{-1}$, almost the same frequency as that for the neutral. On the other hand, all the C=C
stretching modes display frequency shift on oxidation from 0 to +1. The $\nu_a$ mode shows the smallest shift of 32 cm$^{-1}$, from 1566 to 1534
cm$^{-1}$. The $\nu_b$ mode shifts by 84 cm$^{-1}$ but it is losing most of its activity in Raman spectra (see Fig. \ref{fig2}b). The most
charge sensitive and also the strongest $\nu_c$ mode undergoes frequency downshift by 153 cm$^{-1}$ to 1380 cm$^{-1}$. Therefore, we can expect
the $\nu_c$ mode to be the best probe of charge distribution in the EDT-TTF-OX-based organic conductors. Nevertheless, the attribution of the
C=C stretching modes in experimental spectrum of (EDT-TTF-OX)$_2$Mo$_6$Cl$_{14}$ is not straightforward. In particular, the strongest mode that
might be assigned as $\nu_c$ for EDT-TTF-OX$^{+1}$ is found far away from the position of the mode in theoretical spectrum. Here we assign the
1410 cm$^{-1}$ feature as the $\nu_c$ mode of the cation, and 1533 cm$^{-1}$ as $\nu_a$. Let us assume the linear relationship between the
frequency of the $\nu_c$ mode and charge on the molecule. This is true if electron-molecular vibration (EMV) effects are weak (see
\cite{Yamamoto05} and references therein). As a result we get the experimental frequency shift for $\nu_c$ between the neutral and the oxidation
state +1 being 122 cm$^{-1}$. So  that we obtain the relationship between frequency and charge, $\rho$, for the $\nu_c$ mode as $\nu_c$($\rho$)
= 1410 + 122(1-$\rho$). This formula contains a small error related to the fact that Raman spectrum of EDT-TTF-OX$^0$ was measured at room
temperature, and the spectrum of (EDT-TTF-OX)$_2$Mo$_6$Cl$_{14}$ was taken at 25 K; we can expect the $\nu_c$ to appear about 1405 cm$^{-1}$ for
EDT-TTF-OX$^{+1}$ at room temperature. Now, the strong 1469 cm$^{-1}$ feature found in the Raman response of (EDT-TTF-OX)$_2$Mo$_6$Cl$_{14}$ can
be identified as the $\nu_c$ mode of EDT-TTF-OX$^{+0.5}$. We suggest that the appearance of this unusually strong mode is somehow related to the
charge transfer excitation found about 5400 cm$^{-1}$ in the absorption spectrum of (EDT-TTF-OX)$_2$Mo$_6$Cl$_{14}$ (not shown) and peculiar
dimerized structure \cite{Rethore05a}. Contrary to that only donor molecules with charge +1 were found in the structural study of this material
\cite{Rethore05a}, so that more experiments on other 1:1 salts of EDT-TTF-OX would be necessary to clarify this question. In the frequency range
1300 - 1700 cm$^{-1}$, theoretical infrared spectrum of EDT-TTF-OX$^{+1}$ shows weak $\nu_{CN}$ and $\nu_a$ modes together with relatively
strong $\nu_b$ and $\nu_c$ (Fig. \ref{fig3}b). On the other hand, in the experimental infrared spectrum of (EDT-TTF-OX)$_2$Mo$_6$Cl$_{14}$ two
higher frequency features are significantly stronger than the others (Fig. \ref{fig3}b). We attribute the 1637 cm$^{-1}$ mode as $\nu_{CN}$. The
broad and strong feature centered at 1537 cm$^{-1}$ is probably related to the $\nu_a$ mode of EDT-TTF-OX$^{+1}$ and also of
EDT-TTF-OX$^{+0.5}$. The frequency range 1390 - 1460 cm$^{-1}$ probably contains $\nu_b$- and $\nu_c$-related features of both EDT-TTF-OX$^{+1}$
and EDT-TTF-OX$^{+0.5}$. A reliable assignment of the specific mode components is very difficult in case of this infrared spectrum because we
probably observe significant EMV-coupling between at least some of the C=C stretching modes and the charge-transfer electronic excitation
centered at about 5400 cm$^{-1}$. In fact, all the modes are totally symmetric so available for coupling from theoretical point of view, but the
real effect strongly depends on the coupling constants \cite{Girlando11} that are unknown in case of EDT-TTF-OX. The EMV-coupled modes in
infrared spectra are usually strong, broad, and redshifted comparing respective features observed in Raman spectra.\\

\begin{figure}
\includegraphics[width=0.5\textwidth]{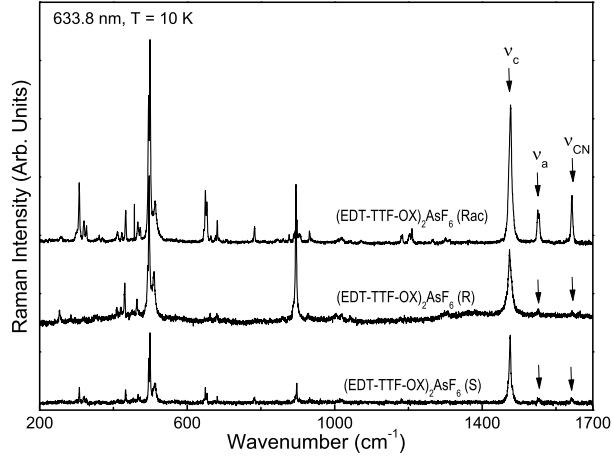}
\caption{Single-crystal Raman spectra of the (EDT-TTF-OX)$_2$AsF$_6$ (Rac), (EDT-TTF-OX)$_2$AsF$_6$ (R) and EDT-TTF-OX)$_2$AsF$_6$ (S) materials
at 10 K. The electrical vector of the laser beam was parallel to the direction of the maximum C=C bands intensity. Note that the spectra are
offset for clarity.\label{fig4}}
\end{figure}

Having done this basic assignment of the C=C stretching modes, we can now analyze results measured for the (EDT-TTF-OX)$_2$AsF$_6$ chiral
organic conductors. Figure \ref{fig4} displays the 10 K Raman spectra of the three materials of interest here. In general, all the materials
have rather similar response, the Raman spectrum of (EDT-TTF-OX)$_2$AsF$_6$ (Rac) being of the best quality. They mostly display Raman-active
vibrational modes of the EDT-TTF-OX donor molecule. The spectra will be discussed in detail in a separate paper. Here we concentrate on the
frequency range of the C=C stretching modes. In order to investigate charge distribution in the whole temperature range, we measured Raman
spectra at several temperatures between 10 K and room temperature, for the racemic (EDT-TTF-OX)$_2$AsF$_6$ (Rac) material and two enantiopure
salts (EDT-TTF-OX)$_2$AsF$_6$ (R) and (EDT-TTF-OX)$_2$AsF$_6$ (S). We looked for both the signatures of different fractional charges and changes
in charge distribution in temperature range of the M-I
phase transition. The results we obtained were similar for all the investigated materials.\\

\begin{figure}
 \includegraphics[width=0.4\textwidth]{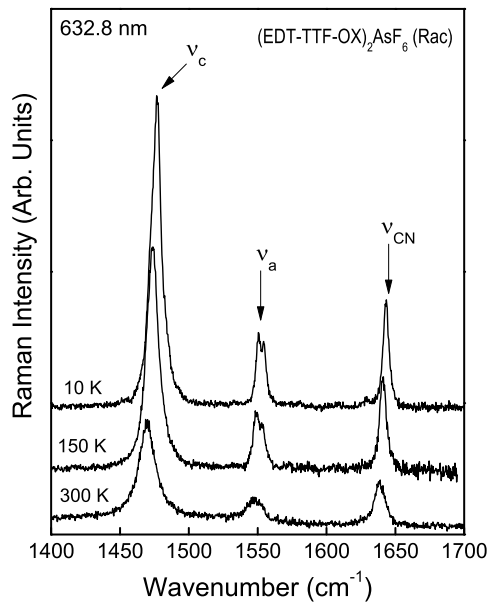}
 \caption{Temperature dependence of the Raman spectra of
 the racemic (EDT-TTF-OX)$_2$AsF$_6$ salt in the frequency range of the C=C stretching modes. Note that the spectra are offset for
 clarity.\label{fig5}}
 \end{figure}

In this paper we discuss the results for the racemic (EDT-TTF-OX)$_2$AsF$_6$  material. Displayed in the Fig. \ref{fig5} are Raman spectra for
selected temperatures 10 K, 150 K and 300 K. First, let us examine the strongest $\nu_c$ feature centered between 1470 (room temperature) and
1477 cm$^{-1}$ (low temperature). In case of charge order at low temperature we could expect a doublet related to two sets of differently
charged EDT-TTF-OX molecules. In the whole temperature range we observe a single band instead, and its frequency confirms uniform charge
distribution +0.5 per EDT-TTF-OX molecule, according to our experimental formula. On the other hand, a doublet structure for the $\nu_a$ mode
 is centered about 1548 cm$^{-1}$ at room temperature. This doublet only slightly changes with lowering the temperature and results in
 two components, 1551 and 1554 cm$^{-1}$ at 10 K. That we do not observe
 any splitting for the most charge sensitive $\nu_c$ at the same time suggests that the doublet is not related with CO but rather with some structural
 disorder.\\

\begin{figure}
\includegraphics[width=0.4\textwidth]{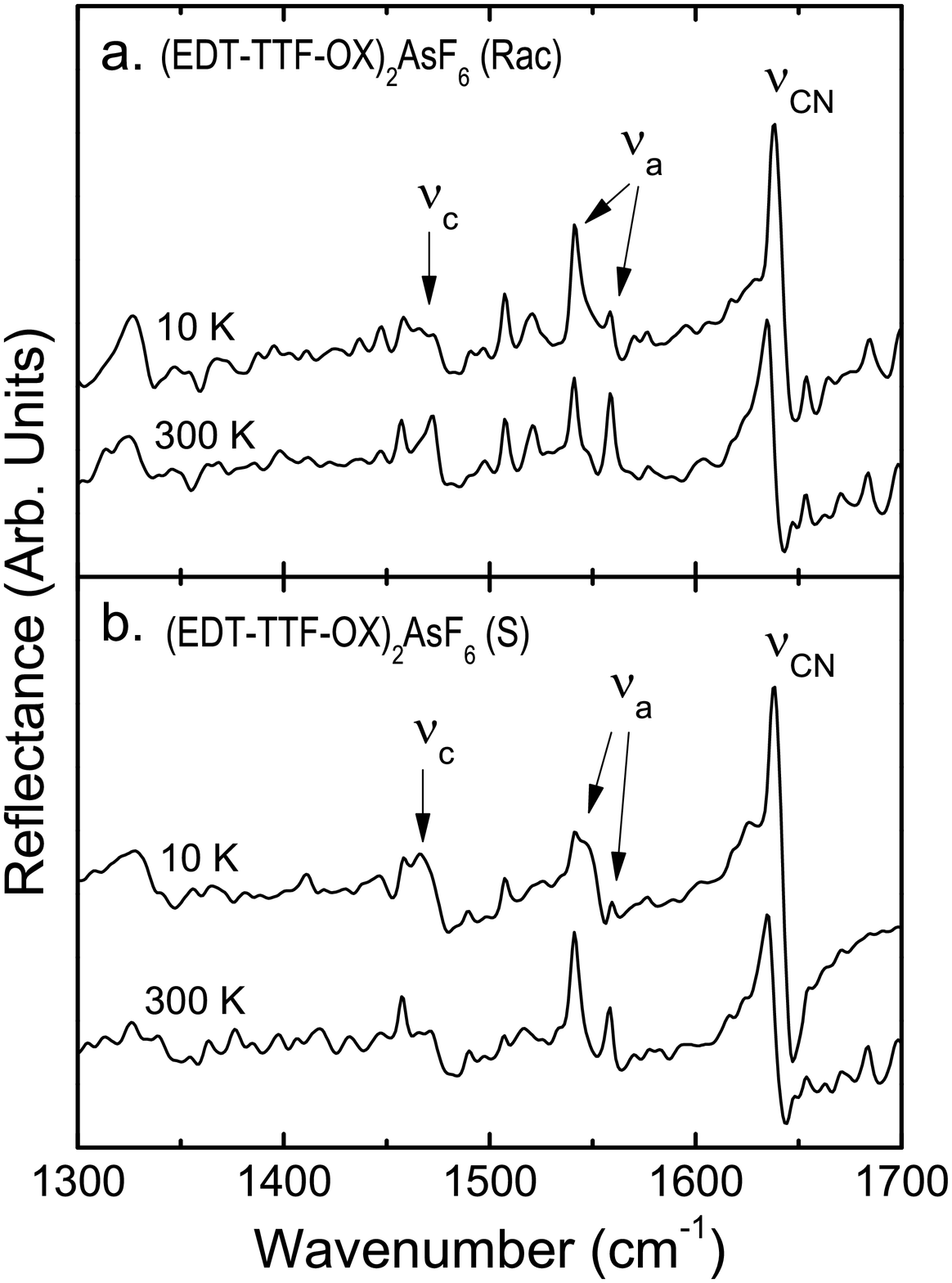}
\caption{Reflectance spectra at 10 and 300 K of (EDT-TTF-OX)$_2$AsF$_6$ (Rac) (a) and (EDT-TTF-OX)$_2$AsF$_6$ (S) (b) polarized along $a$, in
the frequency range of the C=C stretching modes. Note that the spectra are offset for clarity.\label{fig6}}
\end{figure}

 We have also investigated temperature dependence of the polarized reflectance of the
 (EDT-TTF-OX)$_2$AsF$_6$ materials. Figure \ref{fig6} displays close-up view of the reflectance spectra of the racemic (Fig.
 \ref{fig6}a) and enantiopure (S) material (Fig. \ref{fig6}b) at temperatures 10 and 300 K in the low-conducting $a$-direction
 (detailed discussion of the reflectance spectra will be published in a following paper). The doublet structure attributed to the $\nu_a$
 mode is found in all spectra at about 1541 and 1558 cm$^{-1}$. Therefore, the doublet is not related to chirality itself. The frequency difference
 between two mode components is substantially larger in the infrared than in the Raman spectra, the fact that is probably due to EMV-coupling.\\

In summary, we investigated infrared reflectance and Raman spectra of the (EDT-TTF-OX)$_2$AsF$_6$ (Rac), (EDT-TTF-OX)$_2$AsF$_6$ (R) and
EDT-TTF-OX)$_2$AsF$_6$ (S) chiral organic conductors, looking for signatures of charge ordering at low temperature. Using DFT calculations, we
identified the charge-sensitive
$\nu_c$ mode, that can be regarded as a counterpart of the $\nu_{27}$ mode in the BEDT-TTF-based organic conductors \cite{Yamamoto05}.
Our results indicate that charge distribution is uniform in all the investigated materials in the whole temperature range.
Therefore, we can exclude CO as a mechanism responsible for M-I phase transition in this family of materials.\\

The work at the Institute of Molecular Physics Polish Academy of Sciences was supported by the National Science Centre (Decision No.
DEC-2012/04/M/ST3/00774).

\end{document}